\documentclass[prb,twocolumn,showpacs]{revtex4}
\usepackage{graphicx}
\usepackage{amsmath}

\begin{document}
\title{Simulation of a spin-wave instability from atomistic spin dynamics}

\author{J. Hellsvik}
\email[]{johan.hellsvik@fysik.uu.se}
\affiliation{Department of Physics and Materials Science, Uppsala University, Box 530, SE-751 21 Uppsala, Sweden}
\author{B. Skubic}
\affiliation{Department of Physics and Materials Science, Uppsala University, Box 530, SE-751 21 Uppsala, Sweden}
\author{L. Nordstr\"{o}m}
\affiliation{Department of Physics and Materials Science, Uppsala University, Box 530, SE-751 21 Uppsala, Sweden}
\author{O. Eriksson}
\affiliation{Department of Physics and Materials Science, Uppsala University, Box 530, SE-751 21 Uppsala, Sweden}
 
\date{\today}

\begin{abstract}
We study the spin dynamics of a Heisenberg model at finite temperature in the presence of an external field or a uniaxial anisotropy. For the case of the uniaxial anisotropy our simulations show that the macro moment picture breaks down. An effect which we refer to as a spin-wave instability (SWI) results in a non-dissipative Bloch-Bloembergen type relaxation of the macro moment where the size of the macro moment changes, and can even be made to disappear. This relaxation mechanism is studied in detail by means of atomistic spin dynamics simulations.
\end{abstract}
 
\maketitle
\section{Introduction}
Relaxation processes for magnetization dynamics are poorly understood but play a crucial role for spin dynamics in general. In this article we address one of the most fundamental processes of magnetization dynamics, the uniform motion of the magnetization in an anisotropy field. The uniform motion of the magnetization and the relaxation of the uniform motion is of central importance in applications (magnetic switching in storage media etc.).\cite{Stohr06} We study here a ferromagnetic system which initially is excited by a finite angle rotation of the magnetization with respect to the anisotropy axis. This excitation brings the magnetization into a uniform motion where the magnetization eventually relaxes back to an alignment with the magnetization direction parallel to the anisotropy axis. Different phenomenological models, such as Gilbert damping and Bloch-Bloembergen damping,\cite{Bloch46, Bloembergen50} have been used for describing this macro level relaxation of the magnetization. In this article we perform simulations of magnetization dynamics on an atomic scale and we study the consequences for the macro scale behavior of the magnetization dynamics. 

The initial rotation of the magnetization of a ferromagnet in an external field can be seen as an excitation of a large number of uniform k=0 magnons. During the relaxation process these magnons interact, dissipating energy and angular momentum. Relaxation can occur via two processes, one where both energy and angular momentum are transfered out (or in) of the magnetic system and the second where energy is transfered within the magnetic system, to other non-uniform $k\neq0$ magnons. The first process, which describes a dissipative damping in the equations of motion for magnetization dynamics, results in a Gilbert like relaxation, 

\begin{equation}
\frac{\partial \mathbf{M}}{\partial t}=-\gamma \mathbf{M} \times \mathbf{H}+\frac{\alpha}{M} \mathbf{M} \times \frac{\partial \mathbf{M}}{\partial t},
\end{equation}
where $\mathbf{M}$ is the macro moment, $\mathbf{H}$ the effective field, $\gamma$ the gyromagnetic ratio, and $\alpha$ a damping parameter. The second process, which is described by the precessional term in the equations of motion, results in a special case ($|M_z|$ constant) of the Bloch-Bloembergen damping,
\begin{equation}
\frac{\partial \mathbf{M}}{\partial t}=-\gamma \mathbf{M} \times \mathbf{H}-\frac{M_x}{T}\mathbf{\hat{e}}_x-\frac{M_y}{T}\mathbf{\hat{e}}_y,
\end{equation}

where the effective field is assumed to lie in the $z$-direction and $T$ is a relaxation parameter. This second process is the focus of this work. 

We address here a mechanism for the relaxation of the uniform motion of the magnetization within the spin-system itself which is seen to result in a Bloch-Bloembergen like damping of the magnetization of the system. Several such mechanisms exist, such as the Suhl instability,\cite{Suhl57} 2-magnon scattering,\cite{Arias99,Hurben98} 4-magnon scattering,\cite{Dobin03-1} etc. All of the mentioned mechanisms rely on the dipolar interactions resulting in an energy lowering of non-uniform magnons and an energy degeneracy between the uniform magnons and certain non-uniform magnons. 

In this article we study a different mechanism, since dipolar interaction are not even included in our simulations. The mechanism that we address is instead due to the thermal fluctuations on an atomic scale of the magnetization combined with the nature of the uniaxial anisotropy. Such a mechanism, which does not rely on dipolar interactions, were studied by Safonov \textit{et al.} \cite{Safonov01} and recently by Kashuba \cite{Kashuba06} and Garanin \textit{et al.} \cite{Garanin08,Garanin09} where it was shown that a spin wave instability (SWI) develops in a uniaxial anisotropy field. As we show in this article, based on theoretical considerations, the instability should develop on the atomic length scale as well as on the micrometer length scale which was treated by Kashuba, provided certain conditions are fulfilled. The instability is shown to be caused by the altering of the non-uniform thermal magnetic excitations of the system as it undergoes a uniform rotation.

\section{Details of the Spin-dynamics simulations}
There are at least two approaches for studying magnetization dynamics in simulations. Most common is what is pursued in micromagnetics, the solution of the phenomenological Landau Lifshitz Gilbert (LLG) equation on a micrometer length scale for a continuum magnetization.\cite{Aharoni00} An alternative approach, which is utilized here, is based on solving the equations of motion for magnetization dynamics where the magnetization on a nanoparticle or atomic scale is represented by a Heisenberg Hamiltonian. The Heisenberg Hamiltonian is often successful in describing magnetic systems on an atomic scale, especially when using first principles calculations of the interatomic exchange. The current method is based on atomistic spin dynamics which has as starting point a quantum mechanical description from density functional theory of the evolution of the atomic spins. Other works which have taken this approach can be found in Refs.~\onlinecite{Antropov96,Ujfalussy04,Tao05}.

Our simulations are performed using the ASD (Atomic Spin Dynamics) package\cite{Skubic08} which is based on an atomistic approach of spin dynamics. Interatomic exchange and magneto-crystalline anisotropy (MA) are included in the Hamiltonian. We use a parameterization of the interatomic exchange part of the form of a Heisenberg Hamiltonian, where the exchange parameters are calculated from first principles theory. The effect of temperature is modeled by Langevin dynamics. Connection to an external thermal bath is modeled with a Gilbert like damping. The simulations are performed on bcc Fe using four coordination shells in the Heisenberg Hamiltonian. In order to ease comparison we used the same exchange parameters as in Refs.~\onlinecite{Tao05,Skubic08}.

\section{Dynamics in an external field}
\begin{figure}
\includegraphics[width=0.40\textwidth]{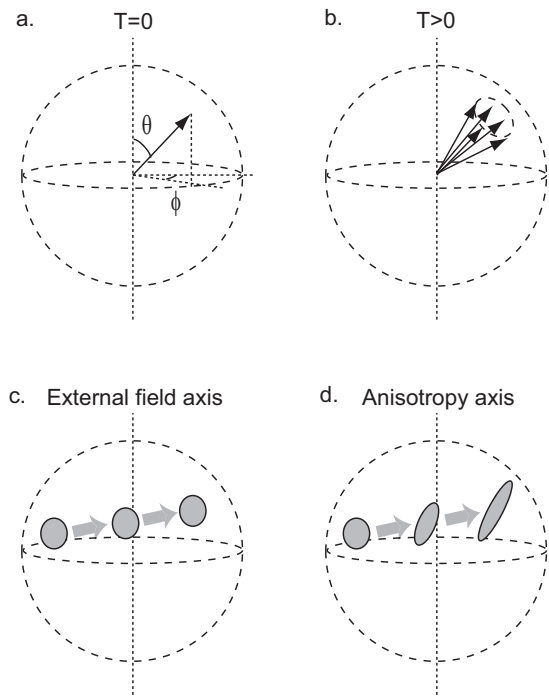}
\caption{\label{fig:macrospin}Figures (a) and (b) show the distribution of atomic moments of the spin dynamics simulations. At finite temperature the orientations of the atomic spins are distributed around a common axis (b). The angles $\theta$ and $\phi$ discussed in the text are also defined (a). Figures (c) and (d) show the evolution of the spin distribution, as given by the evolution of the circular grey disc representing the distribution of magnetic moments defined in (b). The system is at finite temperature in an external field (c) and in a uniaxial anisotropy (d).}
\end{figure}

Different coarse grained levels can be used for describing magnetization dynamics. Here we will work with two levels: (1) the individual atomic moments, $\mathbf{m}_i$, and (2) a macro moment, $\mathbf{M}$, representing the sum of the individual atomic moments of the total system. Any atomic moment is typically exposed to an interatomic exchange field, $\mathbf{B}_{\textrm{eff},i}$, of the order of 1000~T. At finite temperature the atomic moments fluctuate around a common direction. On average, below the critical temperature, there is a finite magnetic moment and the interatomic exchange field averaged over all atoms is directed along the average moment. The size of the average moment or the macro moment depends on the spread of the individual atomic moments, which is governed by the temperature. The situation is illustrated in the top part of Fig.~\ref{fig:macrospin}, where the distribution of atomic moments is illustrated for T=0 K (Fig.~\ref{fig:macrospin}a) and at finite temperature (Fig.~\ref{fig:macrospin}b). This description of magnetization at finite temperature is the starting point for our discussion. 

If the system is exposed to an external field, the average moment will precess in this external field. The atomic moments precess in a uniform way without distortion of their internal distribution. The torque exerted by the external field, $\partial \mathbf{m}_i/\partial t=-\gamma \mathbf{m}_i \times \mathbf{B}_{\textrm{eff},i}$ on each atom $i$, results in an equal angular velocity of all the atomic moments. This is illustrated in Fig.~\ref{fig:model} (top left) where the torque (or $\partial \mathbf{m}_i/\partial t$) is shown as a function of angle ($\theta$) between the magnetic moment and applied field. In Fig.~\ref{fig:model} (top right) we also show the resulting angular velocity ($\partial {\phi}_i/\partial t$) of each atomic spin. The angular velocity is constant and seen to be independent of $\theta$, hence the angular velocity is the same for all spins and it stands clear that an external field will not influence the relative orientation of the atomic spins. The evolution of the distribution of the atomic moments at finite temperature during relaxation in an external field is schematically shown in Fig.~\ref{fig:macrospin}c. The figure illustrates the fact that an external field results in a simple rotation of the magnetization and that all individual atomic spins rotate without changing the relative direction to all other atomic spins.

\section{Dynamics in a uniaxial anisotropy field}
\begin{figure}
\includegraphics[width=0.40\textwidth]{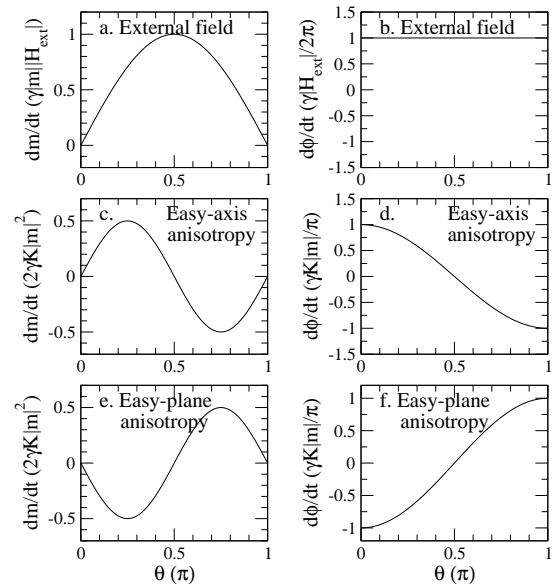}
\caption{\label{fig:model}The plots illustrate the change of the magnetic moment due to an external field, easy-axis anisotropy and an easy-plane anisotropy. The graphs on the left hand side give the magnitude $|\partial\mathbf{m}/\partial t|$ while the graphs on the right hand side give the angular velocity of the atomic spins with respect to angle $\theta$ between spin and applied field or anisotropy axis. Note that in the case of a uniaxial anisotropy field $\theta$ is defined as the angle between moment and a fixed crystallographic direction of the anisotropy field (e.g. 100). H is the strength of the external field and K the strength of the anisotropy field.}
\end{figure}

If there is a uniaxial anisotropy in the system, such as magneto-crystalline anisotropy or shape anisotropy, an excitation of the macro moment in the anisotropy (by a rotation) will in general lead to a precessional motion of the macro moment in the anisotropy field which appears similar to the precession in an external field. For the case of a uniaxial magneto-crystalline anisotropy, which we will consider now, there are however important differences in the spin dynamics. We define the anisotropy energy for each atomic moment as $E=ke_z^2$ where $k$ is the anisotropy constant which determines the strength of the anisotropy and $e_z$ is the $z$-component of the direction of the atomic moment. The torque and angular velocity on any atomic spin are illustrated in Fig.~\ref{fig:model} for both an easy-axis anisotropy (middle panels) and an easy-plane anisotropy (lower panels). The torque is clearly different than in the case of an applied field (top panels), and more importantly the angular velocity of each spin is no longer independent of $\theta$. Hence, spin dynamics in a uniaxial anisotropy field does not lead to a uniform rotation of the atomic spins. Instead the internal distribution of the atomic moments is distorted during the rotation, as illustrated in Fig.~\ref{fig:macrospin}d. For the case of the external field the distribution of atomic spins remained constant with the result that the size of the macro moment remained constant during the precession. Hence, the process could conveniently be described within a macro moment picture. This is not true for the case of the uniaxial anisotropy and the macro moment description breaks down. Since the internal distribution of the atomic moments is changed during the precession, and the direction of any atomic moment in general changes relative to all other moments in the system, the size of the macro moment changes which leads to a considerably more complex macro level behavior. In the rotating frame of the average moment, the easy-axis anisotropy is seen to counteract the precession of atomic moments in the effective exchange field, while the hard-axis anisotropy is seen to enhance the precession in the effective exchange field. As the average moment precesses in the uniaxial anisotropy the atomic moments will have a tendency to spread reducing the net moment of the system, as shown in Fig.~\ref{fig:model}d. We will refer to this behavior as a spin-wave instability (SWI), according to the discussion by Kashuba (Ref.~\onlinecite{Kashuba06}).

\begin{figure}
\includegraphics[width=0.40\textwidth]{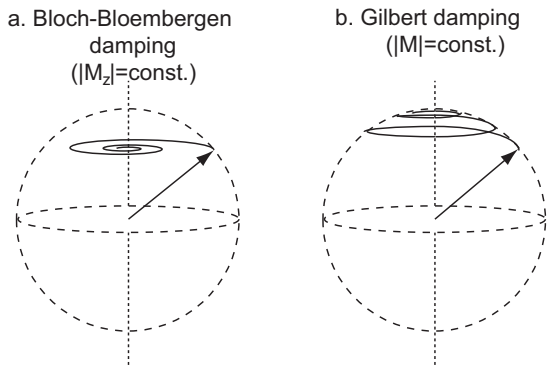}
\caption{\label{fig:BBandG}The figures illustrate the Bloch-Bloembergen (a) damping and the Gilbert (b) damping for a macro moment.}
\end{figure}

As we will show in our simulations, the SWI results in an apparent damping of the uniform motion of the macro moment. We define the anisotropy axis as the $z$-axis. What is significant for this damping is the vanishing of the macro moment components perpendicular $(x,y)$ to the anisotropy axis and the constant value of the parallel macro moment component $(z)$. Hence, during the SWI, the average magnetization of the system drops and only the $z$-component of the average magnetization remains finite as the $x$ and $y$-components vanish. This gives the relaxation of the macro moment due to the SWI a Bloch-Bloembergen form where $|M_z|$ is constant. Thus, due to the SWI there is an alignment of the macro moment with the anisotropy axis where the alignment occurs maintaining a constant value of the $z$-component of the macro moment. This is illustrated in Fig.~\ref{fig:BBandG}. This shows that a redistribution of angular momentum and energy within the magnetic system is taking place. Hence, there is a relaxation taking place even though the dissipative damping, $\alpha$, is set to zero. In reality there is also a finite dissipative damping, $\alpha$, and therefore also a Gilbert contribution to the relaxation of the macro moment. In some of our simulations, in order to clearly observe the Bloch-Bloembergen damping with $M_z$=constant, we set $\alpha=0$. The fact that the value of the $z$-component of the macro moment is constant during the SWI is expected since with zero damping the precessional torque of the uniaxial anisotropy is the only source or drain of angular momentum within the spin system and this torque lacks $z$-component. 

\subsection{Simulating bcc Fe with different strengths of uniaxial anisotropy} 
\begin{figure}
\includegraphics[width=0.40\textwidth]{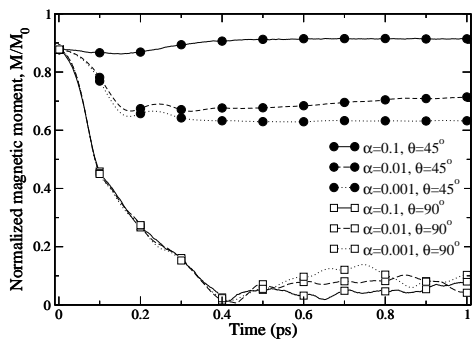}
\caption{\label{fig:damping}Calculated evolution of the total magnetization of bcc Fe as a function of time for different angles between the initial magnetization and the applied field and for different values of the damping parameter. The temperature was 100~K.}
\end{figure}

In order to study the SWI of bcc Fe we choose a 20$\times$20$\times$20 cell with periodic boundary conditions, encompassing 16000 atomic spins, and three different values of the strength of an uniaxial anisotropy: -2~mRy/atom, -0.2~mRy/atom and -0.02~mRy/atom, with an easy axis directed along the $z$-axis. Materials with Fe atoms in a bcc environment and enhanced anisotropy may be found experimentally in magnetic multilayers, e.g. with Pt. The anisotropy can here be significantly stronger than in the bulk case. The magnetic anisotropy of a tetragonal FeCo/Pt(001) superlattice was measured \cite{Warnicke07} to $K_u=2.28$~MJm$^{-3}$, corresponding to $K_u\approx 0.012$~mRy/atom. The perpendicular magnetic anisotropy of (Co, Fe)/Pt multilayers was measured by Sato \textit{et al.} \cite{Sato04-1} to $K_u=0.25$~erg/cm$^{-2}$, corresponding to $K_u\approx 0.027$~mRy/atom. The strongest magnetic anisotropy found in experiments is for SmCo$_5$ \cite{Strnat67,Nesbitt68} with values of $K_u=7.7$~MJm$^{-3}$, corresponding to $K_u\approx 0.31$~mRy/formula unit. As will be presented below we see SWI phenomena in our simulations for the anisotropy values -2~mRy/atom, -0.2~mRy/atom but not for -0.02~mRy/atom. In order to simplify, we have not considered non-magnetic (e.g. Pt) atoms in the simulations, but only the effect they have on the uniaxial anisotropy field. As we will show, these systems can display an instability on a time scale of picoseconds (shown in Fig.~\ref{fig:damping}). We now investigate the dependence of the SWI on thermal fluctuations and damping and we investigate the redistribution of the atomic moments which takes place. 

In Fig.~\ref{fig:damping}, we show a series of simulations for three different damping parameters, $\alpha$, and two different initial angles, $\theta = 45^{\circ}$ and $\theta = 90^{\circ}$. In these simulations we used a uniaxial energy of -2~mRy/atom. For the macro moment there is now a Bloch-Bloembergen like damping due to the SWI and a Gilbert like damping due to the inclusion of a dissipative damping in the microscopic equations of motion. For the case of $\theta=45^{\circ}$ we see the presence of both these damping terms (see Fig.~\ref{fig:damping}). For $\alpha$=0.1 the Gilbert term is seen to dominate. After a short dip in the magnitude of the magnetization due to the SWI the magnitude of the magnetization is seen to recover. For $\alpha$=0.01 and 0.001 the Bloch-Bloembergen damping is seen to dominate, and the size of the magnetic moment reaches a value of M/M$_0 \approx$ 0.6-0.7. For $\theta=90^{\circ}$ the situation is slightly different. At this specific angle only the SWI contributes to the relaxation of the system. For this reason the behavior in Fig.~\ref{fig:damping} is fairly independent of the magnitude of $\alpha$, and the magnetization evolves with time to a value where M/M$_0 \approx$ 0-0.1.

\begin{figure}
\includegraphics[width=0.40\textwidth]{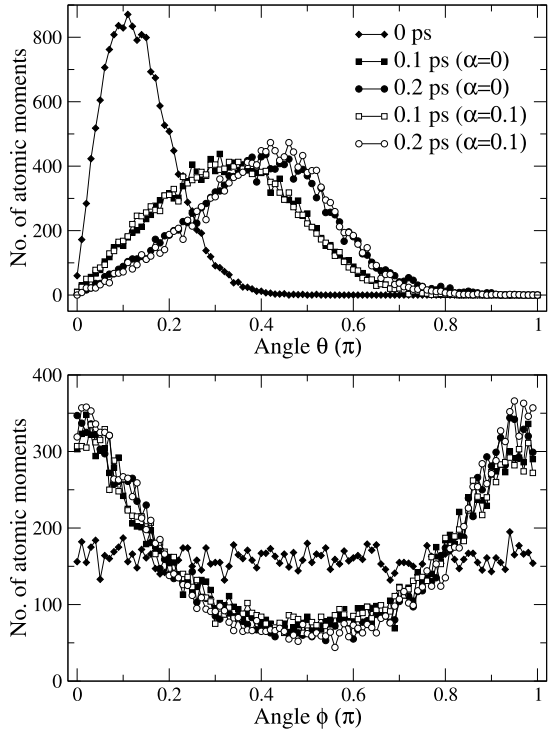}
\caption{\label{fig:hist}Distribution of the angles between the average macro moment and each atomic moment of the bcc Fe simulation cell. In the simulation the initial angle between the average magnetization and the anisotropy axis is $\theta=90^{\circ}$. The top panel shows the distribution of $\theta$ for the different atomic spins and the bottom panel shows the distribution of $\phi$, defined in Fig.~\ref{fig:macrospin}.} 
\end{figure}

The cause of the SWI is an internal redistribution of the atomic moments. In Fig.~\ref{fig:hist} we show a histogram of the angles of the atomic moments with respect to the average atomic moment. The distribution is shown for different points in time for two damping parameters, $\alpha=0.0$ and $\alpha=0.1$ (with a uniaxial energy of -2~mRy/atom). As a first observation, in contrast to what one might expect, the size of $\alpha$ does not change the rate at which the directions of the atomic moments are redistributed. This is illustrated in both the upper and lower panel of Fig.~\ref{fig:hist}. One would, simplemindedly, expect a large damping of the atomic moments in the interatomic exchange field to reduce the spread of the atomic moments, which would counteract the SWI. However, this does not happen. A second observation (see upper panel of Fig.~\ref{fig:hist}) is that the distribution of $\theta$ is smeared out during the SWI. This is consistent with the fact that the net moment of the system is reduced. A third observation (see lower panel of Fig.~\ref{fig:hist}) is that the distribution of $\phi$ is heavily distorted during the SWI. At $t=0$ the distribution is constant, which also is illustrated by the circular disc in Fig.~\ref{fig:macrospin}d. At $t=0.2$~ps the distribution is distorted, which is consistent with the development of an elliptically shaped disc in Fig.~\ref{fig:macrospin}d. 

\begin{figure}
\includegraphics[width=0.40\textwidth]{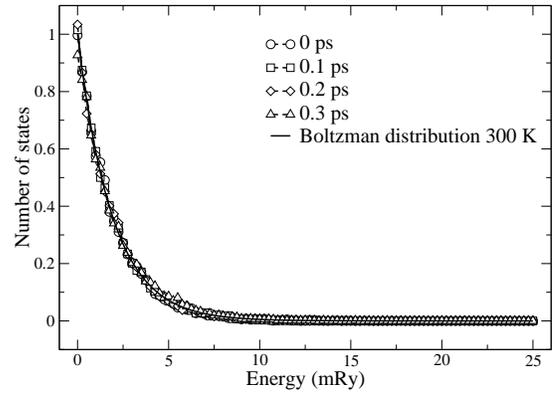}
\caption{\label{fig:dist}Histogram of the energies of the atomic spins for a simulation of bcc Fe  with $\alpha=0.0$ and $\theta=90^{\circ}$. Although there is a large drop of the average moment of the system the energy distribution does not change significantly during the development of the SWI. Data for different times of the simulation are shown.}
\end{figure}

\begin{figure}
\includegraphics[width=0.40\textwidth]{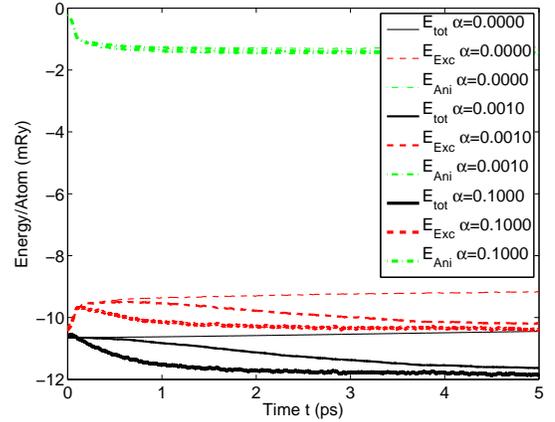}
\caption{\label{fig:energy090a0000}(color online) The evolution in time of the total energy, the exchange energy and the anisotropy energy, for various damping parameters.}
\end{figure}

In order to explain the observations in Fig.~\ref{fig:hist} we show in Fig.~\ref{fig:dist} a histogram of the energy distribution of the magnetic moments at different points in time during the simulation. The histograms for the energy distribution at different points in time fall on top of each other and coincide with the Boltzmann distribution at 300 K, which demonstrates that the simulations are done at thermal equilibrium, throughout the SWI. This explains the first observation of Fig.~\ref{fig:hist}. The effect of the dissipative damping in Langevin dynamics is to bring the system to thermal equilibrium. But since the SWI conserves the thermal distribution of the system, damping has no net effect on the distribution of the directions of the atomic moments. The second and third observation from Fig.~\ref{fig:hist}, concern the change in angular distribution of the atomic moments and explain how the fact that the system remains in thermal equilibrium can be consistent with a reduction in the average magnetization. The angular distribution of the atomic moments is heavily distorted whereas the energy distribution remains constant. For finite damping, the situation changes slightly. We show in Fig.~\ref{fig:energy090a0000} how the magnetic energy, which here is the sum of exchange and anisotropy energy, evolves in time. For the zero damping case in of Fig.~\ref{fig:energy090a0000} the lowering of the anisotropy energy is compensated by an increase in the exchange energy, leaving the total energy constant. This is contrasted by the finite damping cases with $\alpha=0.001$ respective 0.1, were the initial increase in exchange energy decays towards its equilibrium value at the given temperature. In both cases the time-evolution of the atomic moments lowers the total energy. The reason for the different behaviors can be explained as follows.
For the zero damping case, $\alpha=0.000$, the sum of the exchange and the anisotropy energy is a constant of motion. At the start of the simulations the magnetic moments are in thermal equilibrium at $T=300$~K and the total exchange energy is constant. When the anisotropy field is 'turned on' at $t=0$, the system is not in an anisotropy energy minima as the average magnetization is at an angle $\theta=90^{\circ}$ to the easy axis. The evolving magnetic moments lower their anisotropy energy with an amount of energy that is in its entity transfered to exchange energy, since energy can not dissipate in or out of the system. With a small but finite damping, $\alpha=0.001$, the magnetic excitations can dissipate and lower the exchange energy. With a large damping of $\alpha=0.100$ the exchange energy dissipates to within 5 ps to reach its equilibrium value at temperature $T=300$ K.
\begin{figure}
\includegraphics[width=0.40\textwidth]{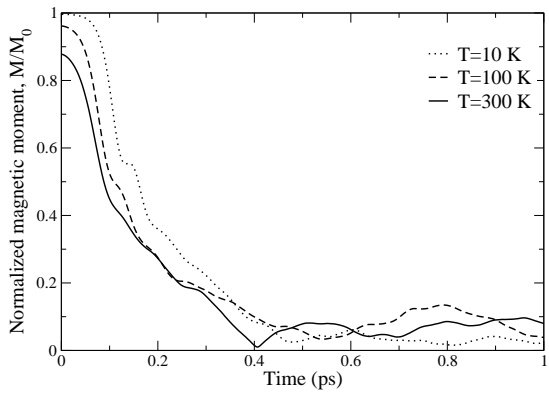}
\caption{\label{fig:temperature}Simulations at different temperatures of bcc Fe with $\alpha=0.0$ and $\theta=90^{\circ}$. The SWI develops on the same time scale for different temperatures.}
\end{figure}

\begin{figure}
\includegraphics[width=0.50\textwidth]{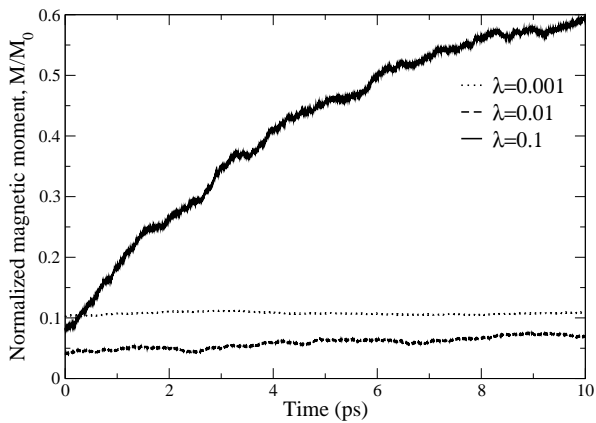}
\caption{\label{fig:relaxback}Starting from bcc Fe in a SWI state the anisotropy field is removed (at T=300 K). The system is seen to evolve back slowly toward a ferromagnetic state.}
\end{figure}

Thermal fluctuations play an important role for the development of the SWI. Naturally there is therefore a dependence of the time-scale of the instability on the temperature. We found however that in the range 10-300 K the time-scale is fairly independent on the temperature as shown in Fig.~\ref{fig:temperature} (again we used a uniaxial energy of -2~mRy/atom for these simulated data). The thermal fluctuations also have another role. For systems where the macro moment is unable to relax along an anisotropy axis (i.e. when $\theta=90^{\circ}$) thermal fluctuations turn out as the only mechanism for the system to come out of the chaotic SWI state when the anisotropy field is removed. Starting from a chaotic state where the SWI has been allowed to bring the system to a zero total moment state we suddenly remove the anisotropy field and observe the evolution of the system (see Fig.~\ref{fig:relaxback}). It is now only the complete randomness of the thermal fluctuations that eventually is able to evolve the system back to a ferromagnetic state. The thermal fluctuations will eventually bring the spin distribution which has a total moment close to zero, to a spin distribution with a total moment approaching a finite value. The process is however very time consuming, as shown in Fig.~\ref{fig:hist}, and only observed for the largest damping parameter in the present simulations. 

\begin{figure}
\includegraphics[width=0.40\textwidth]{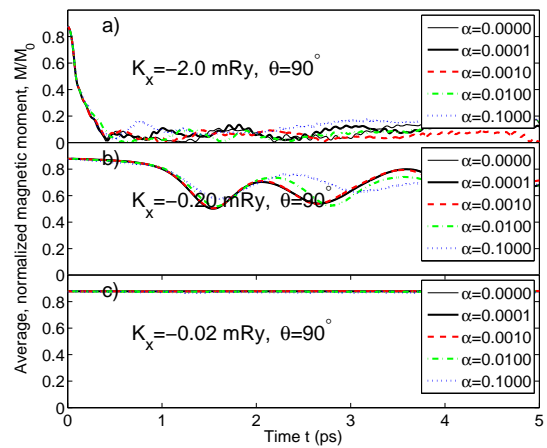}
\caption{\label{fig:theta90}(color online) Calculated evolution of the magnetization of bcc Fe as a function of time with different values of the uniaxial anisotropy and for different values of the damping parameter. The magnetization is initially at angle $\theta = 90^{\circ}$ to the anisotropy axis.}
\end{figure}

We now compare simulated results using different strengths of the uniaxial anisotropy as well as different values of the damping parameter. The interatomic exchange interactions and the size of the simulation cell were kept the same as in previous simulations. The temperature was $300$~K. In Fig.~\ref{fig:theta90} we show the case where $\theta$=90, for three values of uniaxial anisotropy, $-2$~mRy/atom, $-0.2$~mRy/atom and $-0.02$~mRy/atom. Note that the case with an anisotropy of $-2$~mRy/atom was also considered in Fig.~\ref{fig:damping}, although in Fig.~\ref{fig:theta90} we show the dynamical response over a larger time interval, 5 ps. For the strongest value of the uniaxial anisotropy the SWI develops rather easily, whereas for the lowest value of the uniaxial anisotropy the SWI does not develop at all, at least not in the time interval considered. The intermediate value of the uniaxial anisotropy results in an intermediate situation where the macro moment oscillates in time (at least in this time-interval, we will return to this situation below). The reason behind the different behaviors shown in Fig.~\ref{fig:theta90}, is a competition between the strength of the uniaxial anisotropy, which in line with the discussion around Fig.~\ref{fig:macrospin} tends to spread the distribution of all atomic moments, and the importance of the other relevant interactions in the system, primarily the strength of the interatomic exchange interaction.

\begin{figure}
\includegraphics[width=0.40\textwidth]{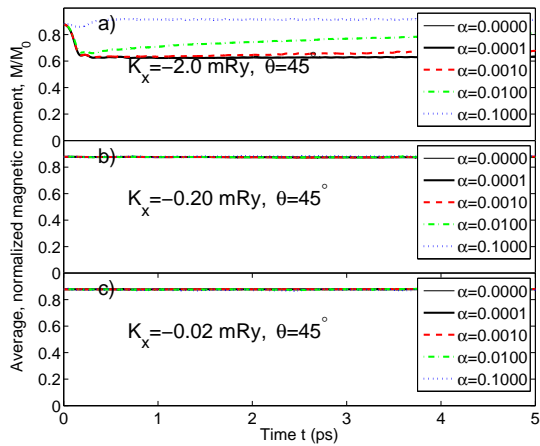}
\caption{\label{fig:theta45}(color online) Same as Fig.~\ref{fig:theta90} but with the magnetization initially at angle $\theta = 45^{\circ}$ to the anisotropy axis.}
\end{figure}

In Fig.~\ref{fig:theta45} we show very similar simulations as in Fig.~\ref{fig:theta90}, with the only difference being that we show results for the case when $\theta=45^{\circ}$. Here the intermediate and lowest value of the uniaxial anisotropy does not have sufficient strength to drive a SWI, whereas the largest value of the anisotropy the SWI develops and a Bloch-Bloembergen damping occurs. This was also illustrated in Fig.~\ref{fig:damping}, but over a shorter time-interval.

\begin{figure}
\includegraphics[width=0.40\textwidth]{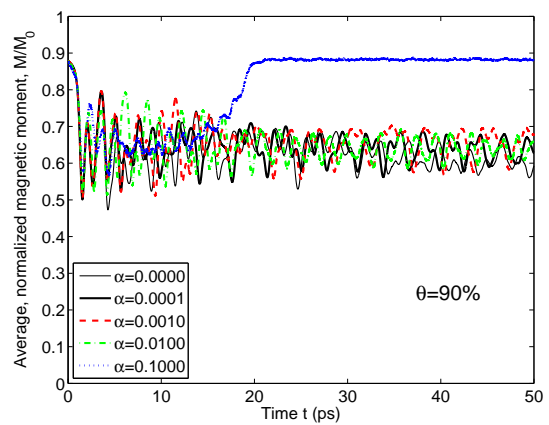}
\caption{\label{fig:theta90B}(color online) Same as the middle panel of Fig.~\ref{fig:theta90} but showing the evolution of the magnetization up to 50 ps. For $\alpha=0\ldots 0.01$ the magnetization oscillates in the interval $0.55-0.70M_0$. For $\alpha=0.1$ the magnetization recovers after $\sim 20$~ps to the value $M=0.88M_0$ which is the thermally equilibrated value at temperature $T=300$~K.}
\end{figure}

The case when $\theta=45^{\circ}$ and with a uniaxial anisotropy of -0.2 mRy/atom is, as Fig.~\ref{fig:theta90} suggests, a particularly interesting case, since here the anisotropy and exchange interactions seems to be tuned into a situation where both are very influential for the evolution of the macro spin. In fact, Fig.~\ref{fig:theta90} suggests that in this case the magnetization oscillates between a Gilbert like damping and Bloch-Bloembergen like damping. For this reason we have extended the simulations over a larger time interval (50 ps), and the results are shown in Fig.~\ref{fig:theta90B}. It is to be noted from this figure that for this borderline case, the evolution of the macro spin depends not only on the competition between interatomic exchange and uniaxial anisotropy, but also on the value of the damping parameter. For large values of the damping a regular Gilbert damping behaviors is found. For small and intermediate values of the damping the macro spin is found to oscillate in time, but otherwise following a dynamic response which resembles Bloch-Bloembergen damping. Hence the data in Fig.~{\ref{fig:theta90B}} show that by careful tuning of the relative importance of the uniaxial anisotropy, exchange interaction and damping, one may obtain a behavior which is more complex than that given by pure Gilbert or Bloch-Bloembergen damping.

The finite size of the simulation cell restricts the possible spin wave excitations. The simulations described so far were all for $L=20$ corresponding to 16000 magnetic moments. With a smaller cell only the modes with short wave lengths can occur. This means that the weaker uniaxial anisotropy cannot drive an SWI unless the simulation cell is large enough. The trend for if a SWI can occur or not for the different simulation cells, with cell size $L=10$, 15, 20 and 25, are presented in Tables~\ref{tab:csize1}, \ref{tab:csize2}, for the 90 degree and 45 degree case, respectively. In the table for the 90 degree case we have defined a strong SWI as the case where the magnetization drops below $0.2M_0$, a medium SWI as when it drops below $0.6M_0$, a weak SWI as when the magnetization drops with $0.05-0.20M_0$ and no SWI when the magnetization drops less than $0.05M_0$. In the table for the 45 degree case the same notation is used apart from that we here redefine strong SWI as when the magnetization drops below $0.65M_0$ (which here corresponds to a Bloch-Bloembergen damping). The results of Tables.~\ref{tab:csize1},\ref{tab:csize2} correspond well to the results of Ref.~\onlinecite{Garanin08} (see e.g. Eqn. 27). For the anisotropy values$-2$~mRy/atom, $-0.2$~mRy/atom and $-0.02$~mRy/atom and with the exchange energy summed up over all coordination shells to $\approx 10$~mRy/atom we get $N_{max}=5,16,51$ where $N_{max}$ is the largest cell size that suppress SWI effects.

\begin{table}
\caption{\label{tab:csize1}Simulations for varying cell size, 90 degree case. The entries describe the possible occurrence of a SWI during the simulation time $t=5$~ps.}
\begin{ruledtabular}
\begin{tabular}{|c|c|c|c|c|c|}
$K_u$ (mRy/atom) & $\alpha$ & L=10 & L=15 & L=20 & L=25 \\
\hline
-2.0 & 0-0.1 & strong & strong & strong & strong\\
-0.2 & 0-0.1 & no & weak & medium & medium \\
-0.02 & 0-0.1 & no & no & no & no \\
\end{tabular}
\end{ruledtabular}
\end{table}

\begin{table}
\caption{\label{tab:csize2}Simulations for varying cell size, 45 degree case. The entries describe the possible occurrence of a SWI during the simulation time $t=5$~ps.}
\begin{ruledtabular}
\begin{tabular}{|c|c|c|c|c|c|}
$K_u$ (mRy/atom) & $\alpha$ & L=10 & L=15 & L=20 & L=25 \\
\hline
-2.0 & 0-0.01 & strong & strong & strong & strong \\
-2.0 & 0.1 & weak & weak & weak & weak \\
-0.2 & 0-0.01 & no & no & no & weak \\
-0.2 & 0.1 & no & no & no & no \\
-0.02 & 0-0.1 & no & no & no & no \\
\end{tabular}
\end{ruledtabular}
\end{table}

\section{Discussion and Summary}
In this paper we have investigated the conditions when a spin-wave instability (SWI) may occur. In order for this to happen, a number of requirements must be met. First, there must be an initial perturbation to the system, e.g. thermal fluctuations, such that the atomic moments start to deviate from the direction of the macro moment. Secondly, there must be a magnetic anisotropy in the system. The presence of a SWI is found to a large degree be determined by a competition between the magnetic anisotropy and the strength of the exchange interaction. In some special cases, where these two contributions are very delicately balanced, the value of the damping parameter can finally determine whether or not a SWI occurs. We have also found that the size of the simulation cell is influential for if a SWI occurs, a conclusion which is in agreement with the results of Ref.\onlinecite{Garanin08}. 

Another conclusion we reach from our simulations is that due to thermal fluctuations the simple model of a macro moment precessing in a uniaxial anisotropy is found to be inaccurate. The uniaxial anisotropy leads to a non-uniform rotation of the composing atomic moments. On a short time scale the effect is small. On a longer time scale or for larger anisotropies there are severe consequences. An instability appears which effectively leads to a Bloch-Bloembergen damping of the magnetization. 

Our simulations point to a technical avenue for designing media for data-storage and magnetic memories, where e.g. the grain size of the storage media would be a materials property which one could compare to the various sizes of our simulation cell. Media with a small grain size could possibly then exhibit a weaker tendency for a SWI to be observed. If experimental evidence for the spin wave instability could be demonstrated, it would imply that there is an increased importance to a fine grain description of the magnetization dynamics in simulations and it would show that macro moment models lose accuracy when anisotropies are involved in the dynamics. Further experimental studies addressing this issue are highly desired. 

\acknowledgments
Financial support from the Swedish Foundation for Strategic Research (SSF), the Swedish Research Council (VR), the Royal Swedish Academy of Sciences (KVA), Liljewalchs resestipendium and Wallenbergstiftelsen is acknowledged. Calculations have been performed at the Swedish national computer centers UPPMAX, HPC2N and NSC.

%\bibliography{/home/johan/fysik4a/Articles/Referencer/JohanReferences.bib}

\begin{thebibliography}{20}
\expandafter\ifx\csname natexlab\endcsname\relax\def\natexlab#1{#1}\fi
\expandafter\ifx\csname bibnamefont\endcsname\relax
  \def\bibnamefont#1{#1}\fi
\expandafter\ifx\csname bibfnamefont\endcsname\relax
  \def\bibfnamefont#1{#1}\fi
\expandafter\ifx\csname citenamefont\endcsname\relax
  \def\citenamefont#1{#1}\fi
\expandafter\ifx\csname url\endcsname\relax
  \def\url#1{\texttt{#1}}\fi
\expandafter\ifx\csname urlprefix\endcsname\relax\def\urlprefix{URL }\fi
\providecommand{\bibinfo}[2]{#2}
\providecommand{\eprint}[2][]{\url{#2}}

\bibitem[{\citenamefont{St\"{o}hr and Siegmann}(2006)}]{Stohr06}
\bibinfo{author}{\bibfnamefont{J.}~\bibnamefont{St\"{o}hr}} \bibnamefont{and}
  \bibinfo{author}{\bibfnamefont{H.-C.} \bibnamefont{Siegmann}},
  \emph{\bibinfo{title}{Magnetism: From fundamentals to nanoscale dynamics}}
  (\bibinfo{publisher}{Springer Verlag}, \bibinfo{address}{Berlin},
  \bibinfo{year}{2006}).

\bibitem[{\citenamefont{Bloch}(1946)}]{Bloch46}
\bibinfo{author}{\bibfnamefont{F.}~\bibnamefont{Bloch}},
  \bibinfo{journal}{Physical Review} \textbf{\bibinfo{volume}{70}},
  \bibinfo{pages}{460} (\bibinfo{year}{1946}).

\bibitem[{\citenamefont{Bloembergen}(1950)}]{Bloembergen50}
\bibinfo{author}{\bibfnamefont{N.}~\bibnamefont{Bloembergen}},
  \bibinfo{journal}{Physical Review} \textbf{\bibinfo{volume}{78}},
  \bibinfo{pages}{572} (\bibinfo{year}{1950}).

\bibitem[{\citenamefont{Suhl}(1957)}]{Suhl57}
\bibinfo{author}{\bibfnamefont{H.}~\bibnamefont{Suhl}},
  \bibinfo{journal}{Journal of Physics and Chemistry of solids}
  \textbf{\bibinfo{volume}{1}}, \bibinfo{pages}{209} (\bibinfo{year}{1957}).

\bibitem[{\citenamefont{Arias and Mills}(1999)}]{Arias99}
\bibinfo{author}{\bibfnamefont{R.}~\bibnamefont{Arias}} \bibnamefont{and}
  \bibinfo{author}{\bibfnamefont{D.~L.} \bibnamefont{Mills}},
  \bibinfo{journal}{Physical Review B} \textbf{\bibinfo{volume}{60}},
  \bibinfo{pages}{7395} (\bibinfo{year}{1999}).

\bibitem[{\citenamefont{Hurben and Patton}(2008)}]{Hurben98}
\bibinfo{author}{\bibfnamefont{M.~J.} \bibnamefont{Hurben}} \bibnamefont{and}
  \bibinfo{author}{\bibfnamefont{C.~E.} \bibnamefont{Patton}},
  \bibinfo{journal}{Journal of Applied Physics} \textbf{\bibinfo{volume}{83}},
  \bibinfo{pages}{4344} (\bibinfo{year}{2008}).

\bibitem[{\citenamefont{Dobin and Victoria}(2003)}]{Dobin03-1}
\bibinfo{author}{\bibfnamefont{A.~Y.} \bibnamefont{Dobin}} \bibnamefont{and}
  \bibinfo{author}{\bibfnamefont{R.~H.} \bibnamefont{Victoria}},
  \bibinfo{journal}{Physical Review Letters} \textbf{\bibinfo{volume}{90}},
  \bibinfo{pages}{167203} (\bibinfo{year}{2003}).

\bibitem[{\citenamefont{Safonov and Neal~Bertram}(2001)}]{Safonov01}
\bibinfo{author}{\bibfnamefont{V.~L.} \bibnamefont{Safonov}} \bibnamefont{and}
  \bibinfo{author}{\bibfnamefont{H.}~\bibnamefont{Neal~Bertram}},
  \bibinfo{journal}{Physical Review B} \textbf{\bibinfo{volume}{63}},
  \bibinfo{pages}{094419} (\bibinfo{year}{2001}).

\bibitem[{\citenamefont{Kashuba}(2006)}]{Kashuba06}
\bibinfo{author}{\bibfnamefont{A.}~\bibnamefont{Kashuba}},
  \bibinfo{journal}{Physical Review Letters} \textbf{\bibinfo{volume}{96}},
  \bibinfo{pages}{047601} (\bibinfo{year}{2006}).

\bibitem[{\citenamefont{Garanin et~al.}(2008)\citenamefont{Garanin, Kachkachi,
  and Reynaud}}]{Garanin08}
\bibinfo{author}{\bibfnamefont{D.~A.} \bibnamefont{Garanin}},
  \bibinfo{author}{\bibfnamefont{H.}~\bibnamefont{Kachkachi}},
  \bibnamefont{and} \bibinfo{author}{\bibfnamefont{L.}~\bibnamefont{Reynaud}},
  \bibinfo{journal}{EPL (Europhysics Letters)} \textbf{\bibinfo{volume}{82}},
  \bibinfo{pages}{17007} (\bibinfo{year}{2008}).

\bibitem[{\citenamefont{Garanin and Kachkachi}(2009)}]{Garanin09}
\bibinfo{author}{\bibfnamefont{D.~G.} \bibnamefont{Garanin}} \bibnamefont{and}
  \bibinfo{author}{\bibfnamefont{H.}~\bibnamefont{Kachkachi}},
  \emph{\bibinfo{title}{Magnetization reversal via internal spin waves in
  magnetic nanoparticles}} (\bibinfo{year}{2009}),
  \bibinfo{note}{http://arxiv.org/abs/0902.1492v1}.

\bibitem[{\citenamefont{Aharoni}(2000)}]{Aharoni00}
\bibinfo{author}{\bibfnamefont{A.}~\bibnamefont{Aharoni}},
  \emph{\bibinfo{title}{Introduction to the theory of ferromagnetism}}
  (\bibinfo{publisher}{Oxford university press}, \bibinfo{address}{Oxford},
  \bibinfo{year}{2000}).

\bibitem[{\citenamefont{Antropov et~al.}(1996)\citenamefont{Antropov,
  Katsnelson, Harmon, Schilfgaarde, and Kusnezov}}]{Antropov96}
\bibinfo{author}{\bibfnamefont{V.~P.} \bibnamefont{Antropov}},
  \bibinfo{author}{\bibfnamefont{M.~I.} \bibnamefont{Katsnelson}},
  \bibinfo{author}{\bibfnamefont{B.~N.} \bibnamefont{Harmon}},
  \bibinfo{author}{\bibfnamefont{M.~v.} \bibnamefont{Schilfgaarde}},
  \bibnamefont{and} \bibinfo{author}{\bibfnamefont{D.}~\bibnamefont{Kusnezov}},
  \bibinfo{journal}{Physical Review B} \textbf{\bibinfo{volume}{54}},
  \bibinfo{pages}{1019} (\bibinfo{year}{1996}).

\bibitem[{\citenamefont{Ujfalussy et~al.}(2004)\citenamefont{Ujfalussy,
  Lazarovits, Szunyogh, Stocks, and Weinberger}}]{Ujfalussy04}
\bibinfo{author}{\bibfnamefont{B.}~\bibnamefont{Ujfalussy}},
  \bibinfo{author}{\bibfnamefont{B.}~\bibnamefont{Lazarovits}},
  \bibinfo{author}{\bibfnamefont{L.}~\bibnamefont{Szunyogh}},
  \bibinfo{author}{\bibfnamefont{G.~M.} \bibnamefont{Stocks}},
  \bibnamefont{and}
  \bibinfo{author}{\bibfnamefont{P.}~\bibnamefont{Weinberger}},
  \bibinfo{journal}{Physical Review B} \textbf{\bibinfo{volume}{70}},
  \bibinfo{pages}{100404} (\bibinfo{year}{2004}).

\bibitem[{\citenamefont{Tao et~al.}(2005)\citenamefont{Tao, Landau, Schulthess,
  and Stocks}}]{Tao05}
\bibinfo{author}{\bibfnamefont{X.}~\bibnamefont{Tao}},
  \bibinfo{author}{\bibfnamefont{D.~P.} \bibnamefont{Landau}},
  \bibinfo{author}{\bibfnamefont{T.~C.} \bibnamefont{Schulthess}},
  \bibnamefont{and} \bibinfo{author}{\bibfnamefont{G.~M.}
  \bibnamefont{Stocks}}, \bibinfo{journal}{Physical Review Letters}
  \textbf{\bibinfo{volume}{95}}, \bibinfo{pages}{087207}
  (\bibinfo{year}{2005}).

\bibitem[{\citenamefont{Skubic et~al.}(2008)\citenamefont{Skubic, Hellsvik,
  Nordstr\"om, and Eriksson}}]{Skubic08}
\bibinfo{author}{\bibfnamefont{B.}~\bibnamefont{Skubic}},
  \bibinfo{author}{\bibfnamefont{J.}~\bibnamefont{Hellsvik}},
  \bibinfo{author}{\bibfnamefont{L.}~\bibnamefont{Nordstr\"om}},
  \bibnamefont{and} \bibinfo{author}{\bibfnamefont{O.}~\bibnamefont{Eriksson}},
  \bibinfo{journal}{Journal of Physics Condensed Matter}
  \textbf{\bibinfo{volume}{20}}, \bibinfo{pages}{315203}
  (\bibinfo{year}{2008}), \urlprefix\url{http://www.fysik.uu.se/cmt/asd/}.

\bibitem[{\citenamefont{Warnicke et~al.}(2007)\citenamefont{Warnicke,
  Andersson, Bj\"{o}rck, Ferr{\'{e}}, and Nordblad}}]{Warnicke07}
\bibinfo{author}{\bibfnamefont{P.}~\bibnamefont{Warnicke}},
  \bibinfo{author}{\bibfnamefont{G.}~\bibnamefont{Andersson}},
  \bibinfo{author}{\bibfnamefont{M.}~\bibnamefont{Bj\"{o}rck}},
  \bibinfo{author}{\bibfnamefont{J.}~\bibnamefont{Ferr{\'{e}}}},
  \bibnamefont{and} \bibinfo{author}{\bibfnamefont{P.}~\bibnamefont{Nordblad}},
  \bibinfo{journal}{Journal of Physics Condensed Matter}
  \textbf{\bibinfo{volume}{19}}, \bibinfo{pages}{226218}
  (\bibinfo{year}{2007}).

\bibitem[{\citenamefont{Sato et~al.}(2004)\citenamefont{Sato, Goto, Ogata,
  Yamaguchi, and Yoshida}}]{Sato04-1}
\bibinfo{author}{\bibfnamefont{T.}~\bibnamefont{Sato}},
  \bibinfo{author}{\bibfnamefont{T.}~\bibnamefont{Goto}},
  \bibinfo{author}{\bibfnamefont{H.}~\bibnamefont{Ogata}},
  \bibinfo{author}{\bibfnamefont{K.}~\bibnamefont{Yamaguchi}},
  \bibnamefont{and} \bibinfo{author}{\bibfnamefont{H.}~\bibnamefont{Yoshida}},
  \bibinfo{journal}{Journal of Magnetism and Magnetic Materials}
  \textbf{\bibinfo{volume}{272-276}}, \bibinfo{pages}{E951}
  (\bibinfo{year}{2004}).

\bibitem[{\citenamefont{Strnat et~al.}(1967)\citenamefont{Strnat, Hoffer,
  Olson, Ostertag, and Becker}}]{Strnat67}
\bibinfo{author}{\bibfnamefont{K.}~\bibnamefont{Strnat}},
  \bibinfo{author}{\bibfnamefont{G.}~\bibnamefont{Hoffer}},
  \bibinfo{author}{\bibfnamefont{J.}~\bibnamefont{Olson}},
  \bibinfo{author}{\bibfnamefont{W.}~\bibnamefont{Ostertag}}, \bibnamefont{and}
  \bibinfo{author}{\bibfnamefont{J.~J.} \bibnamefont{Becker}},
  \bibinfo{journal}{Journal of Applied Physics} \textbf{\bibinfo{volume}{38}},
  \bibinfo{pages}{1001} (\bibinfo{year}{1967}).

\bibitem[{\citenamefont{Nesbitt et~al.}(2009)\citenamefont{Nesbitt, Willens,
  Sherwood, Buehler, and Wernick}}]{Nesbitt68}
\bibinfo{author}{\bibfnamefont{E.~A.} \bibnamefont{Nesbitt}},
  \bibinfo{author}{\bibfnamefont{R.~H.} \bibnamefont{Willens}},
  \bibinfo{author}{\bibfnamefont{R.~C.} \bibnamefont{Sherwood}},
  \bibinfo{author}{\bibfnamefont{E.}~\bibnamefont{Buehler}}, \bibnamefont{and}
  \bibinfo{author}{\bibfnamefont{J.~H.} \bibnamefont{Wernick}},
  \bibinfo{journal}{Applied Physics Letters} \textbf{\bibinfo{volume}{12}},
  \bibinfo{pages}{361} (\bibinfo{year}{2009}).

\end{thebibliography}

\end{document}